\documentclass[usenatbib,usegraphicx]{mn2e}
\usepackage{natbib}
\bibpunct{(}{)}{;}{a}{}{,}
\usepackage{graphicx}
\begin{document}


\title[The expected thermal precursors of GRBs]{The expected thermal precursors of gamma-ray bursts\\ in the internal shock model}
\author[F. Daigne \& R. Mochkovitch]{Fr\'ed\'eric Daigne$^{1,2}$ and Robert Mochkovitch$^{3}$\\
$^{1}$ Max-Planck-Institut f\"ur Astrophysik, Karl-Schwarzschild-Str 1., 81748 Garching bei M\"unchen, Germany\\
$^{2}$ Present address: CEA/DSM/DAPNIA, Service d'Astrophysique, C.E. Saclay, 91191 Gif sur Yvette Cedex, France\\
$^{3}$ Institut d'Astrophysique de Paris, 98 bis bd. Arago 75014 Paris, France\\
}
\maketitle

\begin{abstract}
The prompt emission of gamma-ray bursts probably comes 
from a highly relativistic wind which converts part of its kinetic
energy into radiation via the formation of shocks within
the wind itself. Such "internal shocks" can occur if the wind
is generated with a highly non uniform distribution of the Lorentz
factor.
We estimate the expected photospheric 
 emission of such a
relativistic wind when it becomes transparent. We compare this thermal
emission (temporal profile + spectrum) to the non-thermal
emission produced
by the internal shocks. In most cases, we predict a rather bright
thermal emission 
that should already have been detected. 
This favors
acceleration mechanisms for the wind where the 
initial energy input is under magnetic rather than thermal form.
Such scenarios 
can produce
thermal
X-ray precursors comparable to those observed by GINGA and
WATCH/GRANAT.
\end{abstract}


\section{Introduction}
The cosmological origin of 
long duration gamma-ray
bursts (hereafter GRBs) has been firmly established since
the discovery of their optical counterparts in 1997
\citep{vanparadijs:97}. These late and fading counterparts, the so called
afterglows, have now been detected in many bursts, and in
different spectral ranges\,: X-rays, optical and radio bands.
The redshift 
has been mesured for about 20 GRBs
from $z=0.43$ 
to $z=4.5$. 
The corresponding
isotropic equivalent energy radiated by these GRBs in the gamma-ray range goes
from $5\ 10^{51}\ \mathrm{erg}$ 
to $2\ 10^{54}\ \mathrm{erg}$. 
The beaming factor that has to be taken into account to 
obtain the real amount of radiated energy can be deduced from afterglow observations (achromatic break in the lightcurve, \citet{rhoads:99}). Current estimates lead to a total energy radiated in gamma-rays of about $0.5-1\ 10^{51}\ \mathrm{erg}$ \citep{frail:01}.
The most discussed scenario to explain the GRB phenomenon is
made of three steps :\\
\noindent\textbf{Central engine :} 
The source of GRBs must be able to release a very large amount of energy in
a few seconds. The two most popular candidates are either
the merger of compact objects (neutron star binaries or
neutron star-- black hole systems \citep{narayan:92,mochkovitch:93}) or the gravitational
collapse of a massive star into a black hole
(collapsars/hypernovae \citep{woosley:93,paczynski:98}). 
Such events lead to the formation of very
similar systems made of a stellar mass black hole surrounded
by a thick torus. 
The collapsar
model seems to be favored 
in the case of long bursts 
by observational
evidences that GRBs are located well inside their host galaxy and often associated to
star-forming regions \citep{paczynski:98,djorgovski:01}. 
The released energy  is first injected into an optically thick wind, which
is accelerated via an unknown mechanism,
probably
involving
 MHD processes \citep{thompson:94,meszaros:97,spruit:01} and
becomes eventually relativistic. The existence of such a
relativistic wind has been
directly inferred from the observations of radio
scintillation in GRB 970508 \citep{frail:97} and is also needed to
solve the 
compactness problem and avoid
photon-photon annihilation along the line of sight. Average Lorentz factors larger than 100 are required
\citep{baring:97,lithwick:01}. The next two steps explain how the kinetic energy of
this relativistic wind is converted into radiation at large
distances from the source, when the wind has become optically
thin.\\
\noindent\textbf{Internal shocks :} the production of
gamma-rays is usually associated to the formation of shocks
within the wind itself \citep{rees:94}. Such internal shocks can
appear if the initial distribution of the Lorentz factor is
highly variable, which is very likely considering the
unsteady nature of the envisaged sources \citep{macfadyen:99}. This model has
been studied in details \citep{kobayashi:97,daigne:98,daigne:00}. The main
difficulties which are encountered are a rather low
efficiency for the conversion of the wind kinetic energy into
gamma-rays (a few percents only) and problems in reproducing with synchrotron emission the slope of the low energy part of the spectrum \citep{ghisellini:00}. Despite this
difficulty, the model can successfully reproduce the main
features of the bursts observed by BATSE.\\ 
\noindent\textbf{External shock :} 
the
relativistic wind is decelerated later by the external
medium. This phase of deceleration is probably the best
understood of the three steps and reproduces very well the
afterglow properties \citep{wijers:97}. The dynamics of the wind 
during the deceleration phase
is described by the solution of the relativistic Sedov
problem \citep{blandford:76} and the observed afterglow is due to synchrotron emission
produced by relativistic electrons accelerated behind the
strong forward shock propagating in the external medium
\citep{sari:98}.\\

The work presented in this paper focuses on the 
prompt emission. The spectrum of this 
emission as observed by
BATSE and Beppo-SAX is non-thermal and is well fitted by the
4-parameter ``GRB-function'' proposed by
\citet{band:93}. This function is made of two smoothly
connected power-laws. 
This non-thermal
emission 
probably 
originates from
the radiation of a population of highly
relativistic electrons accelerated behind the shock waves
propagating within the wind during the internal shock phase.\\

Prior to the internal
shock phase, the relativistic wind has to become
transparent. 
At this transition, a thermal emission is produced, that
could contribute to the observed prompt emission.
Parts of the wind can also become opaque at larger radii if internal shocks create pairs in large number. These opaque regions can produce additional thermal components when they become transparent again \citep{meszaros:00}.
Other thermal contributions can be expected, for example when the jet breaks out at the boundary of the stellar envelope in the collapsar scenario \citep{ramirez-ruiz:02}. In this paper, we restrict our analysis to the photospheric thermal component. A similar problem has been studied by \citet{lyutikov:00} in the different context of strongly magnetized winds emitted by rapidly rotating pulsars.\\

The paper is organized as follows :
in sect.~\ref{sec:photosphere} we obtain the position of
the photosphere of a relativistic wind with a highly
variable initial distribution of the Lorentz factor, as
expected in the internal shock model. We then compute the
corresponding photospheric thermal emission in
sect.~\ref{sec:photosphericemission} and compare it to the
non-thermal emission from the internal shocks in
sect.~\ref{sec:Comparison}. The results are discussed in
sect.~\ref{sec:Discussion} and the conclusions are summarized in sect.~\ref{sec:Conclusions}.


\section{The photosphere of a relativistic wind} 
\label{sec:photosphere}
\subsection{Photospheric radius}
\begin{figure}
\resizebox{\hsize}{!}{\includegraphics{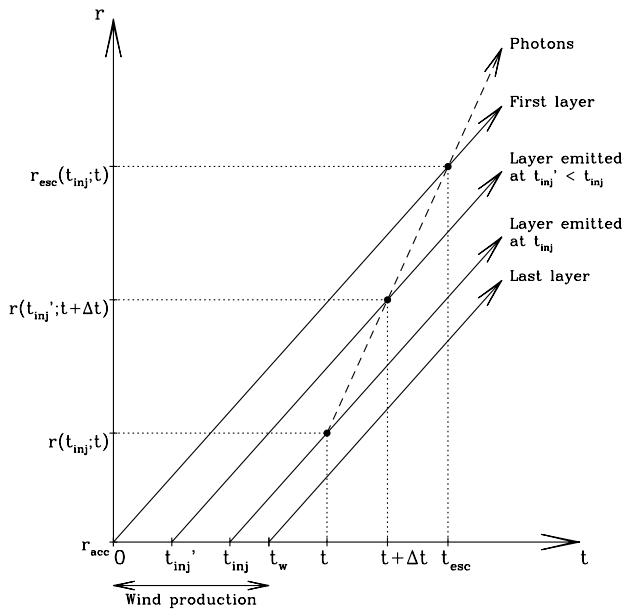}}
\caption{Sketch of the plane $t$-$r$ : the solid line arrows show the
paths of the layers produced by the source at radius $r_\mathrm{acc}$
from $t=0$ to $t=t_\mathrm{w}$. The dashed line arrow shows the path of the
photons emitted at time $t$ by the layer produced at time
$t_\mathrm{inj}$. These photons will cross the layer produced at
time $t_\mathrm{inj}' < t_\mathrm{inj}$ after a duration $\Delta t$. They escape when
they cross the first layer of the wind produced at $t=0$.}
\label{fig:Schema}
\end{figure}
We do not discuss in this paper the nature of the source which is
initially responsible for the energy release leading to the gamma-ray
burst. We suppose that a relativistic wind carrying the energy has
emerged from the source, with an average Lorentz factor $\bar{\Gamma}
\ga 100$. 
We assume that the acceleration is complete at a distance $r_\mathrm{acc}$ from
the source where the ultra-relativistic wind 
is characterized by
an energy
injection rate $\dot{E}(t_\mathrm{inj})$ and an initial distribution of Lorentz
factor $\Gamma(t_\mathrm{inj})$. This corresponds to a mass flux
$\dot{M}(t_\mathrm{inj})=\dot{E}(t_\mathrm{inj}) / \Gamma(t_\mathrm{inj})c^{2}$, 
with $\dot{E}$ and $\dot{M}$ being the isotropic equivalent energy and mass injection rates.
This wind
production process lasts from $t_\mathrm{inj}=0$ to $t_\mathrm{inj}=t_\mathrm{w}$ (all these
quantities are defined in the fixed frame of the source).\\

 In this
section we are interested in computing when the layer emitted by the
source at $t_\mathrm{inj}$ will become transparent. We assume that the wind
is still optically thick at $r_\mathrm{acc}$,
and that it
becomes transparent
before the internal shock phase and before
it is decelerated 
 by the external medium.
We can then consider that each
layer is evolving with a constant Lorentz factor so that at time $t$,
the layer emitted at $t_\mathrm{inj}$ is located at
\begin{equation}
r(t_\mathrm{inj},t) \simeq r_\mathrm{acc}+\left(1-\frac{1}{2\Gamma^{2}(t_\mathrm{inj})}\right) c
(t-t_\mathrm{inj})\ .
\end{equation}
Let us consider photons emitted at $t$ by the layer ejected by the
source at $t_\mathrm{inj}$ (see fig.~\ref{fig:Schema}). 
If they escape from the relativistic wind, these photons will have to cross all the layers emitted from $t_\mathrm{inj}'=0$ to $t_\mathrm{inj}'=t_\mathrm{inj}$.
Precisely, they cross the layer
ejected at $t_\mathrm{inj}'$ after
\begin{equation}
\Delta t \simeq 2 \Gamma^{2}(t_\mathrm{inj}') \frac{\Delta r}{c}\ ,
\label{eq:Cross}
\end{equation}
where $\Delta r$ is the spatial separation at the emission time $t$ between the layer produced
at $t_\mathrm{inj}'$ and the emitting layer produced at time $t_\mathrm{inj}$. This
distance is the initial separation $c\left(t_\mathrm{inj}-t_\mathrm{inj}'\right)$ plus a
correction growing with time due to the difference of Lorentz factor between the two
layers\,:
\begin{eqnarray}
\Delta r & = & r\left(t_\mathrm{inj}',t\right)-r\left(t_\mathrm{inj},t\right) \nonumber\\
& \simeq & 
\left(1-\frac{1}{2\Gamma^{2}\left(t_\mathrm{inj}'\right)}\right) 
c\left(t_\mathrm{inj}-t_\mathrm{inj}'\right)\nonumber\\
& &
+\frac{1}{2}
\frac{ \Gamma^{2}\left(t_\mathrm{inj}'\right) - \Gamma^{2}\left(t_\mathrm{inj}\right)
}{\Gamma^{2}\left(t_\mathrm{inj}'\right)\Gamma^{2}\left(t_\mathrm{inj}\right) }
c\left(t-t_\mathrm{inj}\right)\ .
\label{eq:dr}
\end{eqnarray}
The first term is very close to the initial separation and
the second term is small as long as 
the process we consider takes place well before the internal shock phase.
The photons escape from the wind when they cross the first layer emitted at
$t_\mathrm{inj}'=0$ at time 
\begin{equation}
t_\mathrm{esc} \simeq
t+2\Gamma^{2}(0)\left(t_\mathrm{inj}+\frac{1}{2}\frac{\Gamma^{2}(0)-\Gamma^{2}\left(t_\mathrm{inj}\right)}{\Gamma^{2}(0)\Gamma^{2}\left(t_\mathrm{inj}\right)}\left(t-t_\mathrm{inj}\right)\right)
\end{equation}
and at radius
$\left.r_\mathrm{esc}\right.(t_\mathrm{inj};t) = r(t_\mathrm{inj};t)+c(t_\mathrm{esc}-t)$.
The corresponding distance is $2\Gamma^{2}(0)$ times larger than the
initial separation 
between the emitting layer and the front
of the wind. The total optical depth for these photons is given by 
\begin{equation}
\tau(t_\mathrm{inj},t) = \int_{r(t_\mathrm{inj},t)}^{\left.r_\mathrm{esc}\right.(t_\mathrm{inj},t)}
\mathrm{d}\tau(r)\ .
\end{equation}
The elementary contribution $\mathrm{d}\tau(r)$ to the optical depth
is a Lorentz invariant \citep{abramowicz:91} and is more easily estimated in the comoving
frame of the layer crossed by the photons at $r$ :
\begin{equation}
\mathrm{d}\tau(r) = \kappa \rho' \mathrm{d}l'(r)\ ,
\end{equation}
where $\kappa$ and $\rho'$ are the opacity and the comoving density of
the layer. The length $\mathrm{d}l'(r)$ is computed by a Lorentz
transformation from the fixed frame to the comoving frame of the
layer. We take into account the fact that when photons cover a
distance $dr$, the corresponding duration is $\mathrm{d}t=\mathrm{d}r/c$ so that
\begin{equation}
\mathrm{d}l'(r) = \Gamma \left(\mathrm{d}r - v \mathrm{d}t\right) \simeq \Gamma
\left[1-\left(1-\frac{1}{2\Gamma^{2}}\right)\right] \mathrm{d}r \simeq
\frac{\mathrm{d}r}{2\Gamma}\ .
\end{equation}
The comoving density is given by 
\begin{equation}
\rho' \simeq \frac{1}{\Gamma}\frac{\dot{M}}{4\pi r^{2} c}\ .
\end{equation}
Here all the physical quantities like $\Gamma$, $\dot{M}$, etc. have
the value corresponding to the layer crossed by the photons at $r$,
i.e. the layer emitted at $t_\mathrm{inj}'$ solution of (from eq.~\ref{eq:Cross})
\begin{equation}
r = r(t_\mathrm{inj},t)+2\Gamma^{2}(t_\mathrm{inj}')\Delta r\ ,
\label{eq:rtinj}
\end{equation}
where $\Delta r$ is given by eq.~\ref{eq:dr}.
The final expression for the total optical depth is 
\begin{equation}
\tau(t_\mathrm{inj},t) \simeq \int_{r(t_\mathrm{inj},t)}^{\left.r_\mathrm{esc}\right.(t_\mathrm{inj},t)}
\frac{\kappa\dot{M}}{8\pi \Gamma^{2} r^{2} c} \mathrm{d}r\ .
\label{eq:TauEuler}
\end{equation}
We define the photospheric radius $\left.r_\mathrm{ph}\right.(t_\mathrm{inj})$ of the layer emitted
at $t_\mathrm{inj}$ by
\begin{equation}
\left.r_\mathrm{ph}\right.(t_\mathrm{inj}) = r(t_\mathrm{inj},t)\ \mathrm{with}\ \tau(t_\mathrm{inj},t) = 1\ .
\end{equation}
To estimate this radius, we still need to specify the
opacity. We consider here the phase when the acceleration is
complete. The internal energy has already been almost entirely
converted into kinetic energy. Pairs have annihilated and do not
contribute to the opacity. Then the optical depth is due to the
ambient electrons and the opacity is given by the Thomson opacity
$\kappa=\kappa_\mathrm{T}$. In the following, when a numerical value
is needed, we use $\kappa=0.2$ (i.e a number of electrons per nucleon $Y_{e}=0.5$).
\subsection{The case of a constant Lorentz factor}
\begin{figure}
\resizebox{\hsize}{!}{\includegraphics{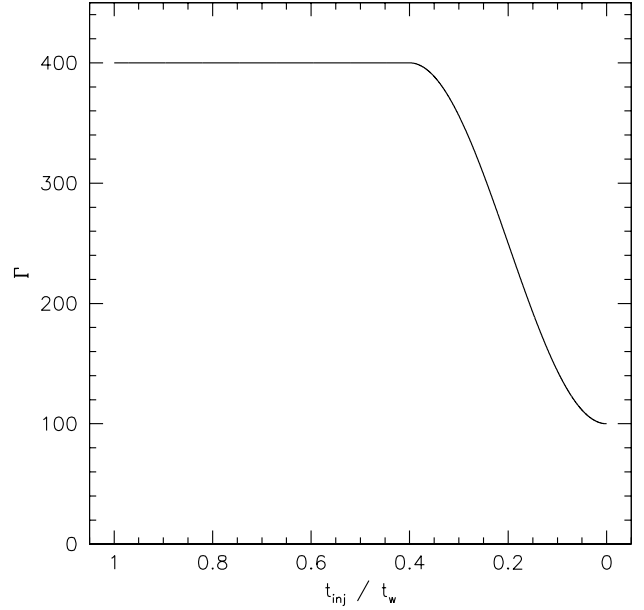}}
\caption{\textbf{A single pulse burst\,: initial distribution of the Lorentz factor}
(example considered in sec.~\ref{sec:Example}).
The Lorentz factor $\Gamma$ is
plotted as a function of the injection time $t_\mathrm{inj}$. As a constant
energy injection rate $\dot{E}$ is assumed, the masses of the fast
($\Gamma=400$)
and ``slow'' parts ($\Gamma=100
\to 400$) are equal.}
\label{fig:GammaInitial}
\end{figure}
In the case of an homogeneous wind where $\dot{E}$, $\dot{M}$ and
$\Gamma$ are constant, we have
\begin{equation}
\tau(t_\mathrm{inj},t) \simeq
\frac{\kappa\dot{M}t_\mathrm{inj}}{4\pi
r(t_\mathrm{inj},t)\left.r_\mathrm{esc}\right.(t_\mathrm{inj},t)}\ .
\end{equation}
There are two limiting cases : 
\begin{equation}
\tau(t_\mathrm{inj},t) \simeq \frac{\kappa\dot{M}t_\mathrm{inj}}{4\pi
r^{2}(t_\mathrm{inj},t)}\ \mathrm{if}\ r(t_\mathrm{inj},t) \gg 2\Gamma^{2}c t_\mathrm{inj}
\end{equation}
or
\begin{equation}
\tau(t_\mathrm{inj},t) \simeq \frac{\kappa\dot{M}}{8\pi c \Gamma^{2}
r(t_\mathrm{inj},t)}\ \mathrm{if}\ r(t_\mathrm{inj},t) \ll 2\Gamma^{2}ct_\mathrm{inj}\ .
\end{equation}
The corresponding photospheric radius of the layer ejected at $t_\mathrm{inj}$ is given
by
\begin{equation}
\left.r_\mathrm{ph}\right.(t_\mathrm{inj}) \simeq \left(\frac{\kappa \dot{M}
t_\mathrm{inj}}{4\pi}\right)^{1/2}\ \mathrm{if}\ \frac{\dot{E}}{\Gamma^{5}t_\mathrm{inj}} \gg \frac{16\pi c^{4}}{\kappa}
\end{equation}
or
\begin{equation}
\left.r_\mathrm{ph}\right.(t_\mathrm{inj}) \simeq \frac{\kappa \dot{M}}{8\pi c \Gamma^{2}}\ \mathrm{if}\ \frac{\dot{E}}{\Gamma^{5}t_\mathrm{inj}} \ll \frac{16\pi c^{4}}{\kappa}\ ,
\label{eq:TauHomo}
\end{equation}
where we have replaced $\dot{M}$ by $\dot{E}/\Gamma c^{2}$. The
condition $\dot{E} / \Gamma^{5}t_\mathrm{inj}  \ll 16\pi
c^{4} / \kappa$ reads
\begin{equation}
\frac{\dot{E}_{52}}{\Gamma_{2}^{5}t_\mathrm{inj}} \ll \frac{200}{\kappa_{0.2}}\ ,
\end{equation}
which is usually true. Here $\dot{E}_{52}$, $\Gamma_{2}$ and
$\kappa_{0.2}$ are respectively $\dot{E}$, $\Gamma$ and $\kappa$ in
unit of $10^{52}\ \mathrm{erg/s}$, $10^{2}$ and $0.2$. Then, the photospheric radius is the same for all
the layers and is given by (eq.~\ref{eq:TauHomo}) :
\begin{equation}
\left.r_\mathrm{ph}\right. = 3.0\ 10^{12} \kappa_{0.2} \dot{E}_{52} \Gamma_{2}^{-3}\
\mathrm{cm}\ .
\end{equation}
If we estimate $r_\mathrm{acc}$ by the saturation radius $\Gamma r_{0}$
which is predicted in the fireball model, we get 
\begin{equation}
r_\mathrm{acc} \simeq 9\ 10^{8}\ \mu_{1}\Gamma_{2}\ \mathrm{cm}
\label{eq:racc}
\end{equation}
 for a typical initial radius $r_{0}$
taken to be the last stable orbit at three Schwarzschild radii around
a non rotating black hole of mass $M_\mathrm{BH}=10 \mu_{1} \mathrm{M}_{\sun}$. It is clear
that $\left.r_\mathrm{ph}\right.$ is much larger than $r_\mathrm{acc}$
as expected.
\subsection{The case of a variable Lorentz factor}
We now consider the case where the initial distribution of the
Lorentz factor is variable. We use the simple model that has
been developed by \citet{daigne:98}. The wind is made of a collection
of ``solid'' layers ejected regularly on a time scale $\Delta t_\mathrm{inj}$ with a
Lorentz factor, a mass and an energy $\Gamma_{i}$,
$M_{i}=\dot{M}_{i}\Delta t_\mathrm{inj}$ and $E_{i}=\dot{E}_{i}\Delta t_\mathrm{inj}$
where $i=1$ corresponds to the first layer 
produced at $t_\mathrm{inj}=0$. Photons emitted by the layer $i_0$ when it is located at
$r_{\left.i_0\right.}$ travel through a total optical depth
\begin{equation}
\tau(r_{\left.i_0\right.}) = \frac{\kappa}{8\pi c^{3}}\sum_{i\le i_0}
\frac{\dot{E}_{i}}{\Gamma_{i}^{3}}\left(\frac{1}{r^{\mathrm{in}}_{i}}-\frac{1}{r^{\mathrm{out}}_{i}}\right)\ ,
\label{eq:exact}
\end{equation}
where $r^{\mathrm{in}}_{i}$ and $r^{\mathrm{out}}_{i}$ are the radii
at which the photons enter and escape the layer $i$. 
We have (from eq.~\ref{eq:rtinj}) 
\begin{equation}
r^{\mathrm{in}}_{i} \simeq
2\Gamma_{i}^{2}\left[c\left(i_0-i\right)\Delta t_\mathrm{inj}+\frac{r_{\left.i_0\right.}}{2\Gamma^{2}_{\left.i_0\right.}}\right]
\end{equation}
and
\begin{equation}
r^{\mathrm{out}}_{i} \simeq r^{\mathrm{in}}_{i}+2\Gamma_{i}^{2}c\Delta t_\mathrm{inj}\ .
\end{equation}
In the following, we
use the exact formula (\ref{eq:exact}) to compute the optical depth and we solve
numerically $\tau(\left.r_\mathrm{ph}\right._{\left.i_0\right.})=1$ to get the photospheric
radius of the layer $i_0$. 
An approximate value is obtained under the assumption that the opacity is dominated by the contribution of the layer $i_0$ where photons are emitted. We then have\,:
\begin{equation}
\left.r_\mathrm{ph}\right._{\left.i_0\right.}^{\mathrm{approx}} \simeq
\frac{\kappa\dot{E}_{\left.i_0\right.}}{8\pi
c^{3}\Gamma^{3}_{\left.i_0\right.}}\ ,
\label{eq:rphapp}
\end{equation}
for $\left.r_\mathrm{ph}\right._{\left. i_0\right.} \ll 2 \Gamma_{\left. i_0 \right.}^{2} c t_\mathrm{inj}$ (with $t_\mathrm{inj}=(i_{0}-1)\Delta t_\mathrm{inj}$). This is usually true except for the first layers ($t_\mathrm{inj} \to 0$).

\begin{figure*}
\resizebox{0.45\hsize}{!}{\includegraphics{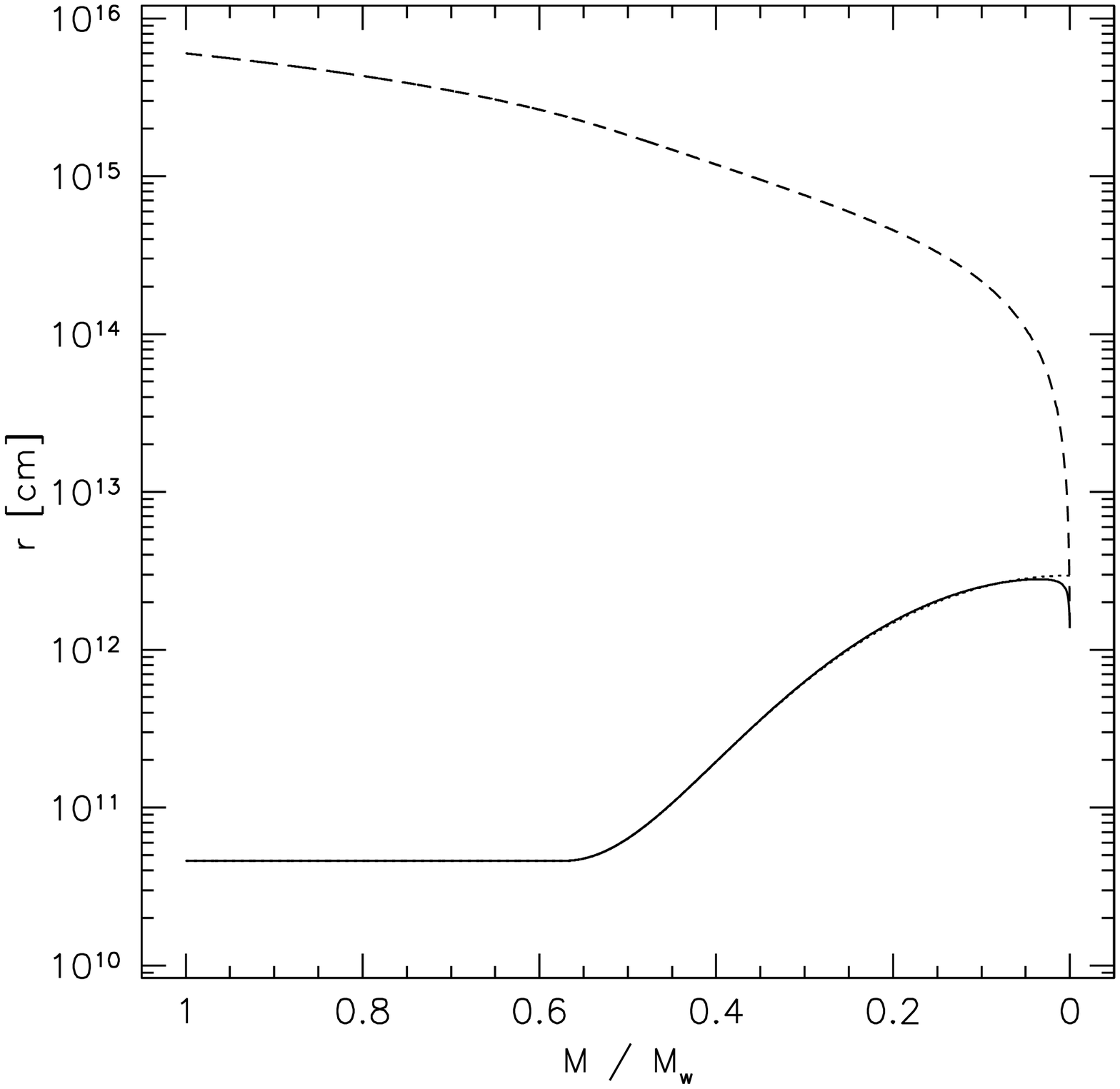}}
\resizebox{0.45\hsize}{!}{\includegraphics{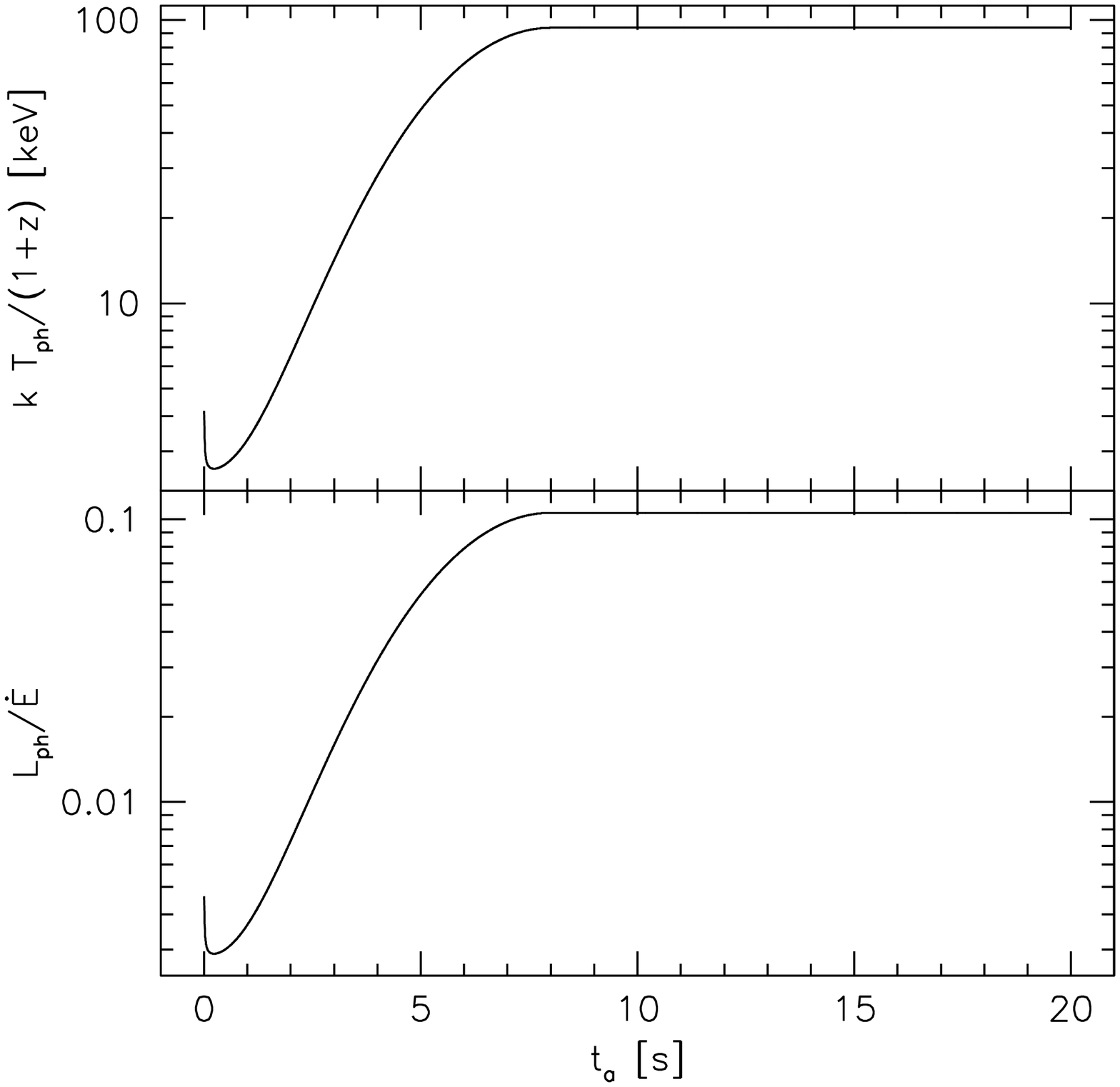}}
\caption{\textbf{A single pulse burst\,: photosphere of the relativistic wind} 
(example considered in secs.~\ref{sec:Example} and \ref{sec:ExamplePh}).
\textbf{Left\,:} the photospheric radius $\left.r_\mathrm{ph}\right.$ is plotted as a function of the
mass coordinate of layers within the relativistic wind (solid
line). The approximate value eq.~\ref{eq:rphapp} (dotted line) is also shown : the only
noticeable difference is at the front of the wind ($M \to 0$). The
dashed line shows the radius at which the photons emitted at the
photospheric radius of each layer escape from the wind. \textbf{Right\,:}
the photospheric luminosity $\left.L_\mathrm{ph}\right.$ and observed temperature  $\left.k T_\mathrm{ph}\right./(1+z)$ 
are plotted as a
function of the arrival time $t_\mathrm{a}$ for $\lambda=1$. A redshift $z=1$ is assumed.}
\label{fig:LTph}
\label{fig:rph}
\end{figure*}
\subsection{Example : a single pulse burst}
\label{sec:Example}
We consider the case of a relativistic wind ejected during $t_\mathrm{w}=10\
\mathrm{s}$ with a constant energy injection rate $\dot{E}=10^{52}\
\mathrm{erg/s}$ and an initial distribution of Lorentz factor represented
in fig.~\ref{fig:GammaInitial}. Such a simple initial distribution has
already been
considered in \citet{daigne:98,daigne:00} and leads to a typical
single pulse burst. 
We use $\Delta t_\mathrm{inj}=2\ 10^{-3}\ \mathrm{s}$
so that the wind is made of $5000$ layers. For each layer $i$ we compute
the photospheric radius $\left.r_\mathrm{ph}\right._{i}$ and the radius $\left.r_\mathrm{esc}\right._i$ where the photons
emitted at the photosphere escape from the relativistic wind. The
result is plotted in fig~\ref{fig:rph} as a function of the mass
coordinate $M_i$. Notice that except for the front of the wind, the
approximate value of $\left.r_\mathrm{ph}\right.$ given by eq.~\ref{eq:rphapp} works
extremely well. An interesting result is that the deepest layers in
the wind become transparent before the layers located at the
front. This is due to the fact that photons emitted by these layers
cross the front at larger radii when the density has already strongly
decreased (one can see that $\left.r_\mathrm{esc}\right.$ for these layers is larger than at
the front).\\
The photospheric radius goes from $\simeq 4.7\ 10^{10}\ \mathrm{cm}$
 to $\simeq 3.0\ 10^{12}\ \mathrm{cm}$. We can check now that this is
well before the internal shocks form or the deceleration of the wind
by the external medium becomes efficient. The typical radius of the
internal shocks is given by
\begin{equation}
r_\mathrm{IS} \simeq f\ \bar{\Gamma}^{2} ct_\mathrm{var}\ ,
\end{equation}
where $t_\mathrm{var}$ is the characteristic time scale for the variations of
the Lorentz factor and $f$ is a numerical factor depending on the
details of the initial distribution of the Lorentz factor ($f$ will be
smaller for high contrasts of $\Gamma$). For a typical average Lorentz
factor $\bar{\Gamma}\ga 100$ we have 
\begin{equation}
r_\mathrm{IS} \simeq 3\ 10^{14}\ f\ \bar{\Gamma}_{2}^{2} t_\mathrm{var}\ \mathrm{cm}
\end{equation}
and we immediately see that except for very small values of $f$ 
or very short time scales $t_\mathrm{var}$,
 the typical radius of the internal shocks 
 is 
larger than the photospheric radius. The deceleration of the wind by the external medium occurs even further away, except in very dense wind environments.


\section{Time profile and spectrum of the photospheric
emission}
\label{sec:photosphericemission}

\subsection{Photospheric luminosity}
In the framework of the fireball model, the temperature and
luminosity of a layer at its photospheric radius $\left.r_\mathrm{ph}\right.$ are given (in
the fixed frame) by (see e.g. \citet{piran:99})
\begin{equation}
\left.k T_\mathrm{ph}\right. \simeq k T^{0} \left(\frac{\left.r_\mathrm{ph}\right.}{r_\mathrm{acc}}\right)^{-2/3}
\label{eq:Tph}
\end{equation}
and
\begin{equation}
\left.L_\mathrm{ph}\right. \simeq \dot{E}
\left(\frac{\left.r_\mathrm{ph}\right.}{r_\mathrm{acc}}\right)^{-2/3}
\label{eq:Lph}
\end{equation}
(in this section we omit the index $i$. Everything applies to each layer). The radius $r_\mathrm{acc}$ is the saturation radius defined by eq.~\ref{eq:racc} and
the initial blackbody temperature of the layer is \citep{meszaros:00}
\begin{equation}
k T^{0} \simeq 1.3\ \dot{E}_{52}^{1/4}\ \mu_{1}^{-1/2}\ \mathrm{MeV}.
\end{equation}
Deviations from the predictions of the standard fireball model are however possible.
The central engine of gamma-ray bursts
is still poorly understood and the acceleration mechanism
not clearly identified. A large fraction of the
energy released by the source may be for instance initially stored under magnetic form
\citep{spruit:01}. In this case, the wind is not as hot as in the
standard fireball model and the photospheric luminosity is also
smaller. An extreme case would be the magnetic acceleration of a cold wind where the photospheric temperature and luminosity 
are negligible.\\

\noindent Whatever the physics of this early phase may be, it should necessarily have the two following properties in common with the standard fireball model\,:
\begin{itemize}
\item[(i)] \textit{The acceleration mechanism must have a good efficiency.} The observed isotropic equivalent gamma-ray luminosity $\left.L_{\gamma}\right.$ is indeed very high. To account for it, the internal shock model requires a isotropic equivalent kinetic energy flux
\begin{equation}
\dot{E}_{52} = \frac{\left.L_{\gamma}\right._{51}}{\left.f_{\gamma}\right._{0.1}}\ ,
\end{equation}
where $\left.L_{\gamma}\right._{51}$ is the observed isotropic equivalent gamma-ray luminosity in unit of
$10^{51}\ \mathrm{erg/s}$ and $\left.f_{\gamma}\right._{0.1}$
is the efficiency for the conversion of kinetic energy into
gamma-rays in unit of $0.1$. As this kinetic energy flux is already very high, we cannot expect the source to release much more energy. Therefore the need for an efficient acceleration is unavoidable. This means that beyond $r_\mathrm{acc}$, the energy flux is completely dominated by the kinetic energy flux, like beyond the saturation radius in the standard fireball model. The main difference may probably be the value of $r_\mathrm{acc}$ compared to the standard saturation radius $\sim \Gamma r_{0}$.
\item[(ii)] \textit{Beyond $r_\mathrm{acc}$ the wind experiences a phase of adiabatic cooling due to spherical expansion.} An efficient acceleration indeed implies that the wind is still optically thick at $r_\mathrm{acc}$. In this case $\left.k T_\mathrm{ph}\right.$ and $\left.L_\mathrm{ph}\right.$ decrease as $r^{-2/3}$ beyond $r_\mathrm{acc}$ like in the sandard fireball model. Only the initial value of the temperature and the internal energy density at the end of the acceleration phase can be different from those of the standard fireball model. 
\end{itemize}
To account for our poor knowledge of the physical process responsible for the acceleration of the wind, 
we
define $\lambda$
as the fraction of the energy which is initially injected under internal energy form. In the standard fireball model $\lambda=1$ whereas $0 \le \lambda \le 1$ in the other possible cases. With this definition we have
\begin{equation}
\left.k T_\mathrm{ph}\right. \simeq \lambda^{1/4} k T^{0} \left(\frac{\left.r_\mathrm{ph}\right.}{r_\mathrm{acc}}\right)^{-2/3}
\label{eq:Tphfint}
\end{equation}
and
\begin{equation}
\left.L_\mathrm{ph}\right. \simeq \lambda \dot{E}
\left(\frac{\left.r_\mathrm{ph}\right.}{r_\mathrm{acc}}\right)^{-2/3}\ .
\label{eq:Lphfint}
\end{equation}
The acceleration radius $r_\mathrm{acc}$
may differ from
the saturation radius given by eq.~\ref{eq:racc}. However,
we will show below that the relevant quantity to
estimate the photospheric emission is the ratio $\left.L_\mathrm{ph}\right./\left.k T_\mathrm{ph}\right.$
which does not depend on $r_\mathrm{acc}$, as long as the photospheric
radius is large compared to $r_\mathrm{acc}$.

\subsection{Spectrum, count rate and arrival time of the
photospheric emission}
\label{sec:ArrivalTime}
\begin{figure*}
\resizebox{0.45\hsize}{!}{\includegraphics{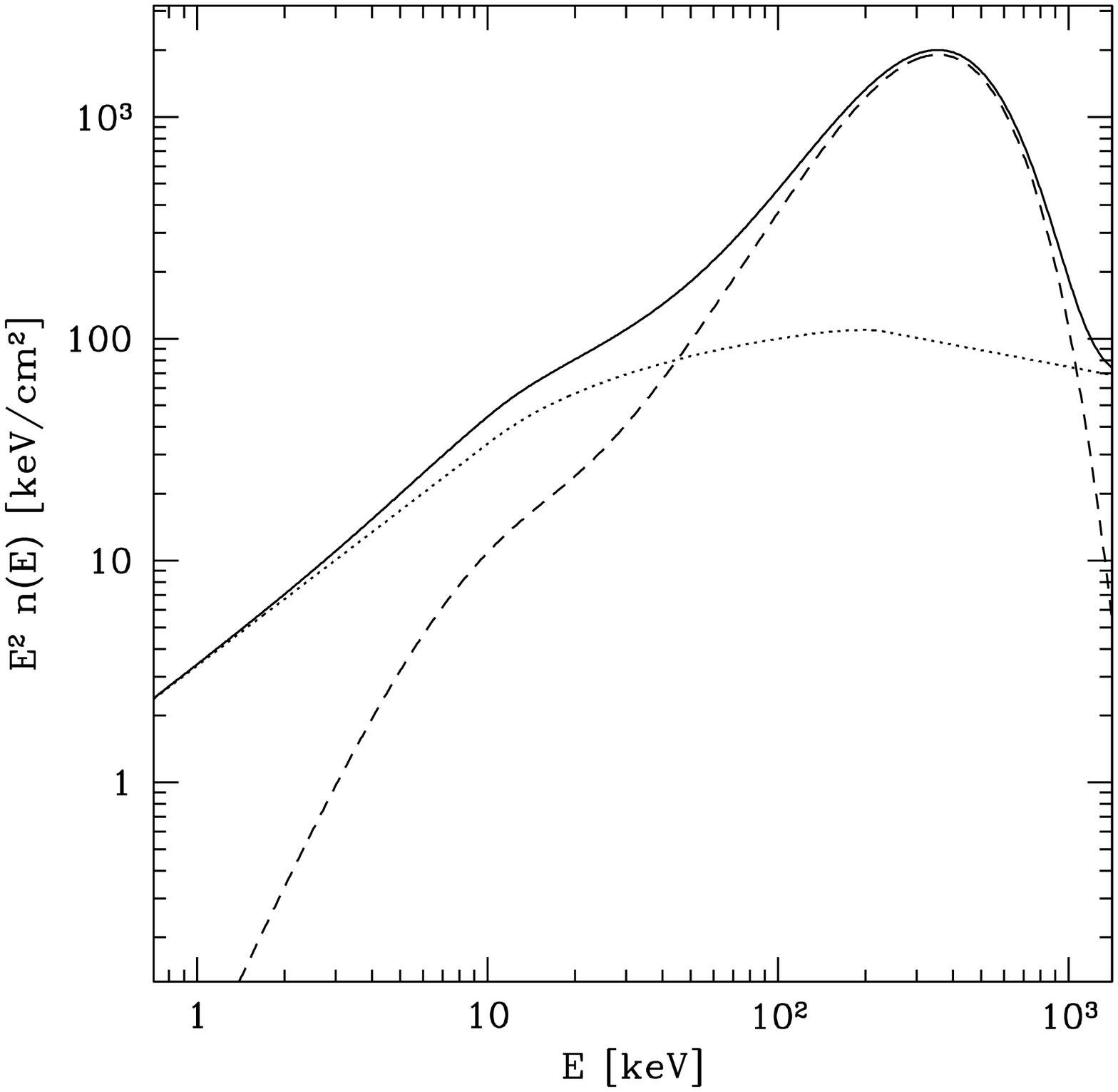}}
\resizebox{0.45\hsize}{!}{\includegraphics{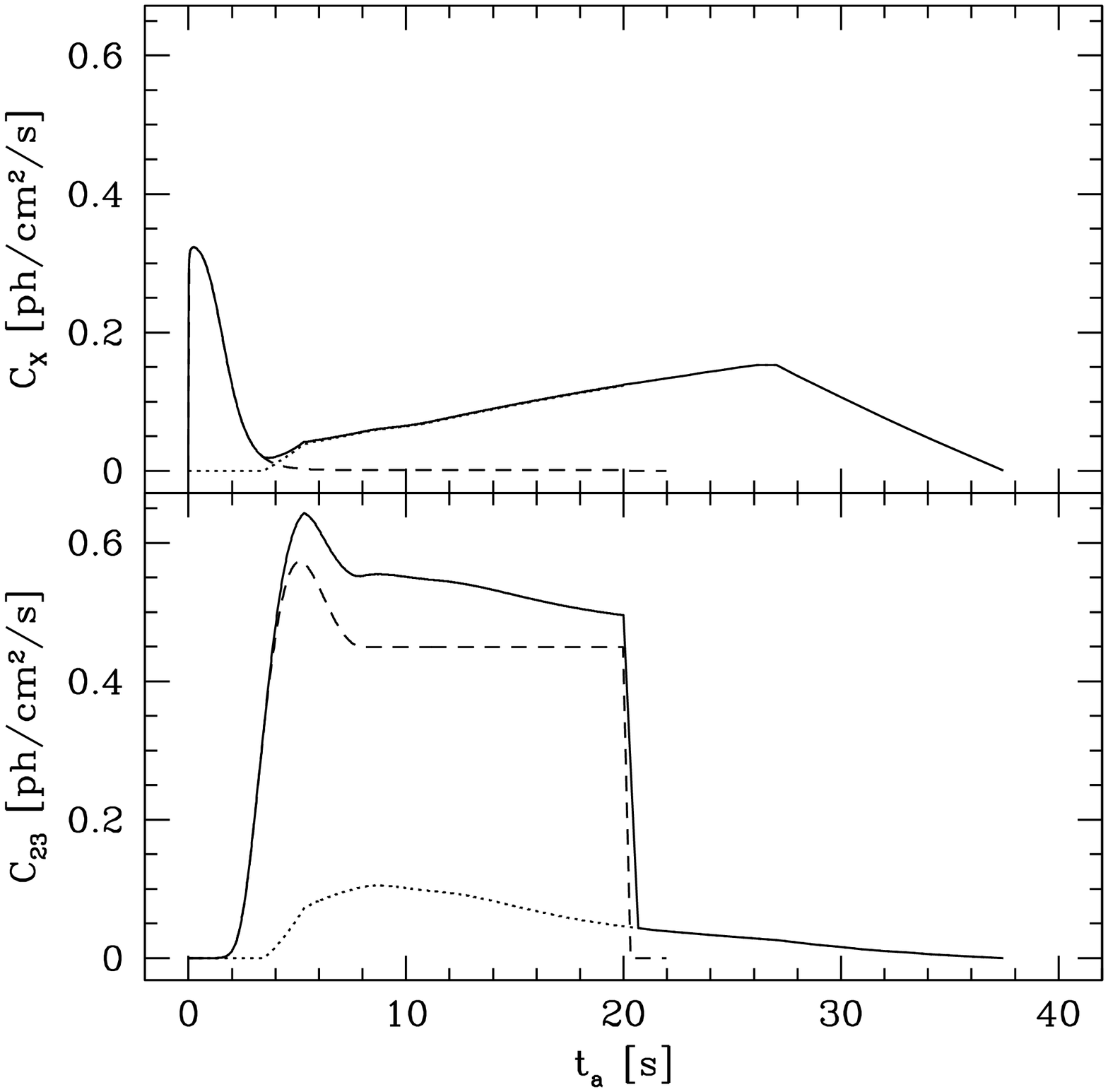}}
\caption{
\textbf{A single pulse burst\,: spectrum and time profile} 
(example considered in  secs.~\ref{sec:Example} and \ref{sec:ExamplePh}).
A redshift $z=1$ is assumed.
\textbf{Left\,:} the spectrum
$E^{2}n(E)$ is plotted as a function of the photon energy in
$\mathrm{keV}$. 
\textbf{Right\,:} the count rate is plotted as a function of arrival time in the $3.5$--$8.5\ \mathrm{keV}$ band (top) and in the $50$--$300\ \mathrm{keV}$ band (bottom).
The dashed line corresponds to the photospheric emission, the dotted line to the non-thermal emission from the internal shocks and the solid line to the total emission.
The global spectrum is
dominated by the luminous contribution from the photosphere of the
rapid part of the wind, peaking at about $3.92 \left.k T_\mathrm{ph}\right. \simeq 370\
\mathrm{keV}$.
}
\label{fig:SpectrumProfiles}
\end{figure*}
We suppose that the photosphere radiates as a blackbody at temperature $\left.k T_\mathrm{ph}\right.$.
This is clearly a simplifying assumption as scattering processes may play an important role when the opacity is $\tau \ga 1$. However, we believe that the possible deviations from a pure blackbody will not change our main conclusions. 
We also neglect corrections in the spectrum due to angular effects affecting photons originating from different regions of the emitting shell. 
We then consider that
the emitted photons have a Planck distribution which is in the source frame :
\begin{equation}
\frac{\mathrm{d}n^{\mathrm{ph}}(E)}{\mathrm{d}E\mathrm{d}t} = \frac{1}{\mathcal{I}_{\mathrm{Planck}}}\ \frac{\left.L_\mathrm{ph}\right.}{\left(\left.k T_\mathrm{ph}\right.\right)^{4}}\ \frac{E^{2}}{\exp{\left(\frac{E}{\left.k T_\mathrm{ph}\right.}\right)}-1}\ ,
\end{equation}
where $\mathcal{I}_{\mathrm{Planck}} = \int_{0}^{+\infty}\frac{x^{3}}{\exp{x}-1}\ dx = \frac{\pi^{4}}{15}$.
Taking into account the redshift $z$ of the source, the observer will detect a photon flux at energy $E$ (observer frame) which is given by
\begin{equation}
C^{\mathrm{ph}}(E) = \frac{1}{\mathcal{I}_{\mathrm{Planck}}}\ \frac{\left.L_\mathrm{ph}\right.}{4\pi D_{\mathrm{L}}^{2}}\ \left(\frac{1+z}{\left.k T_\mathrm{ph}\right.}\right)^{4}  \frac{E^{2}}{\exp{\left(\frac{(1+z)E}{\left.k T_\mathrm{ph}\right.}\right)}-1}\ ,
\end{equation}
where $D_{\mathrm{L}}$ is the luminosity distance at redshift $z$. 
The corresponding count rate in the energy band $\left[E_{1};E_{2}\right]$ is
\begin{equation}
C^{\mathrm{ph}}_{12} = \frac{\left.L_\mathrm{ph}\right.}{4\pi D_{\mathrm{L}}^{2}}\ \frac{1+z}{\left.k T_\mathrm{ph}\right.}\ \frac{\mathcal{I}^{\mathrm{ph}}_{12}}{\mathcal{I}_{\mathrm{Planck}}}\ ,
\end{equation}
where $\mathcal{I}^{\mathrm{ph}}_{12}=\int_{x_{1}}^{x_{2}}\frac{x^{2}}{\exp{x}-1}dx$ and $x_{1,2}=(1+z) E_{1,2}/\left.k T_\mathrm{ph}\right.$. It is interesting to notice that the ratio $\left.L_\mathrm{ph}\right./\left.k T_\mathrm{ph}\right.$ depends neither on the shell radius nor on the saturation radius :
\begin{equation}
\frac{\left.L_\mathrm{ph}\right.}{\left.k T_\mathrm{ph}\right.} \simeq 5.0\ 10^{57}\ \lambda^{3/4}\ \dot{E}_{52}^{3/4}
\mu_{1}^{1/2}\ \mathrm{ph/s}.
\end{equation}
Then the count rate in the energy band $\left[E_{1};E_{2}\right]$ is given by
\begin{equation}
C_{12}^{\mathrm{ph}} \simeq 4.0\ \frac{1+z}{D^{2}_{28}}\ \lambda^{3/4}\ \dot{E}_{52}^{3/4}\ \mu_{1}^{1/2}\ \frac{\mathcal{I}^{\mathrm{ph}}_{12}}{\mathcal{I}_{\mathrm{Planck}}}\ \mathrm{ph/cm^{2}/s}\ ,
\end{equation}
where $D_{28}$ is the luminosity distance $D_{\mathrm{L}}$ in unit of $10^{28}\ \mathrm{cm}$.
The emitted photons will be detected at the arrival time
$t_\mathrm{a}$ (relatively to a signal travelling at the speed of
light)\,:
\begin{equation}
t_\mathrm{a} = \left.t_\mathrm{ph}\right.-\frac{\left.r_\mathrm{ph}\right.}{c}\ ,
\end{equation}
where $\left.t_\mathrm{ph}\right.$ is the time when the layer reaches the radius $\left.r_\mathrm{ph}\right.$. We
get
\begin{equation}
t_\mathrm{a} \simeq t_\mathrm{inj}+\frac{\left.r_\mathrm{ph}\right.}{2\Gamma^{2}c}\ .
\end{equation}
With the approximate value of $\left.r_\mathrm{ph}\right.$ given by eq.~\ref{eq:rphapp}, we
have
\begin{equation}
t_\mathrm{a}^{\mathrm{approx}} \simeq t_\mathrm{inj}+\frac{\kappa\dot{E}}{8\pi
c^{4}\Gamma^{5}}\ .
\label{eq:taPh}
\end{equation}
We 
already checked in the previous section that the second term
is negligible compared to $t_\mathrm{inj}$. Then $t_\mathrm{a} \simeq t_\mathrm{inj}$. The
spreading of arrival times over a duration 
$\Delta t_\mathrm{a} \simeq \left.r_\mathrm{ph}\right. / 2c\Gamma^{2}$ due to the curvature of the
emitting surface is also negligible for the
same reason (of course these estimations of $t_\mathrm{a}$ have to be multiplied
by $1+z$ in the observer frame to 
account for the redshift).
The fact that $t_\mathrm{a} \simeq (1+z)t_\mathrm{inj}$ shows that the time
profile of the photospheric emission, if observed, would provide 
a detailed direct information about 
the initial distribution of Lorentz factor in
the wind. 
\subsection{Example : a single pulse burst}
\label{sec:ExamplePh}
We consider the same distribution of the Lorentz factor and injected power as in sec.~\ref{sec:Example} and we
now compute
the thermal emission of the photosphere for a standard fireball ($\lambda=1$). Fig.~\ref{fig:LTph} shows the 
luminosity and the temperature
at the photosphere as a function of the arrival time of photons. We did not use the approximations given by Eqs.~\ref{eq:Tph}--\ref{eq:Lph} which are strictly valid only for $\left.r_\mathrm{ph}\right. \gg r_\mathrm{acc}$ but we used the exact solution of the fireball equations (see e.g. \citet{piran:99}).
We adopted a redshift $z=1$.
Fig.~\ref{fig:SpectrumProfiles} shows the corresponding integrated spectrum of the photospheric
emission
and the time profile 
in two energy bands\,: $3.5$--$8.5\
\mathrm{keV}$  which is one of the X-ray bands of \textit{Beppo-SAX}
and $50$--$300\ \mathrm{keV}$ which is the 2+3 gamma-ray
band of \textit{BATSE}. The photospheric emission of
 the ``slow'' part ($t_\mathrm{inj}=0 \to 4\ \mathrm{s}$ and $t_\mathrm{a} \simeq
 (1+z)t_\mathrm{inj}=0 \to 8\ \mathrm{s}$) 
has a temperature $\left.k T_\mathrm{ph}\right.$
 increasing from $4.2$ to $94\ \mathrm{keV}$. It
 initially produces a
 pulse only visible in the X-ray band ($t_\mathrm{a}\simeq 0 \to 3\
 \mathrm{s}$). Then, the count rate rises in the gamma-ray band,
 reaches a maximum at $\sim 5\ \mathrm{s}$ when $\left.k T_\mathrm{ph}\right. \simeq 48\ \mathrm{keV}$ and
 starts to decrease (although the temperature is still increasing)
 because the peak energy ($\sim 3.92\ \left.k T_\mathrm{ph}\right.$) becomes larger
 than $300\ \mathrm{keV}$. The rapid part ($t_\mathrm{inj}=4 \to 10\ \mathrm{s}$ and $t_\mathrm{a} \simeq
 (1+z)t_\mathrm{inj}=8 \to 20\ \mathrm{s}$) 
has a constant
 temperature of $94\ \mathrm{keV}$, so that the count rate is constant and mainly visible in the 
 gamma-ray range.
\begin{figure*}
\resizebox{0.45\hsize}{!}{\includegraphics{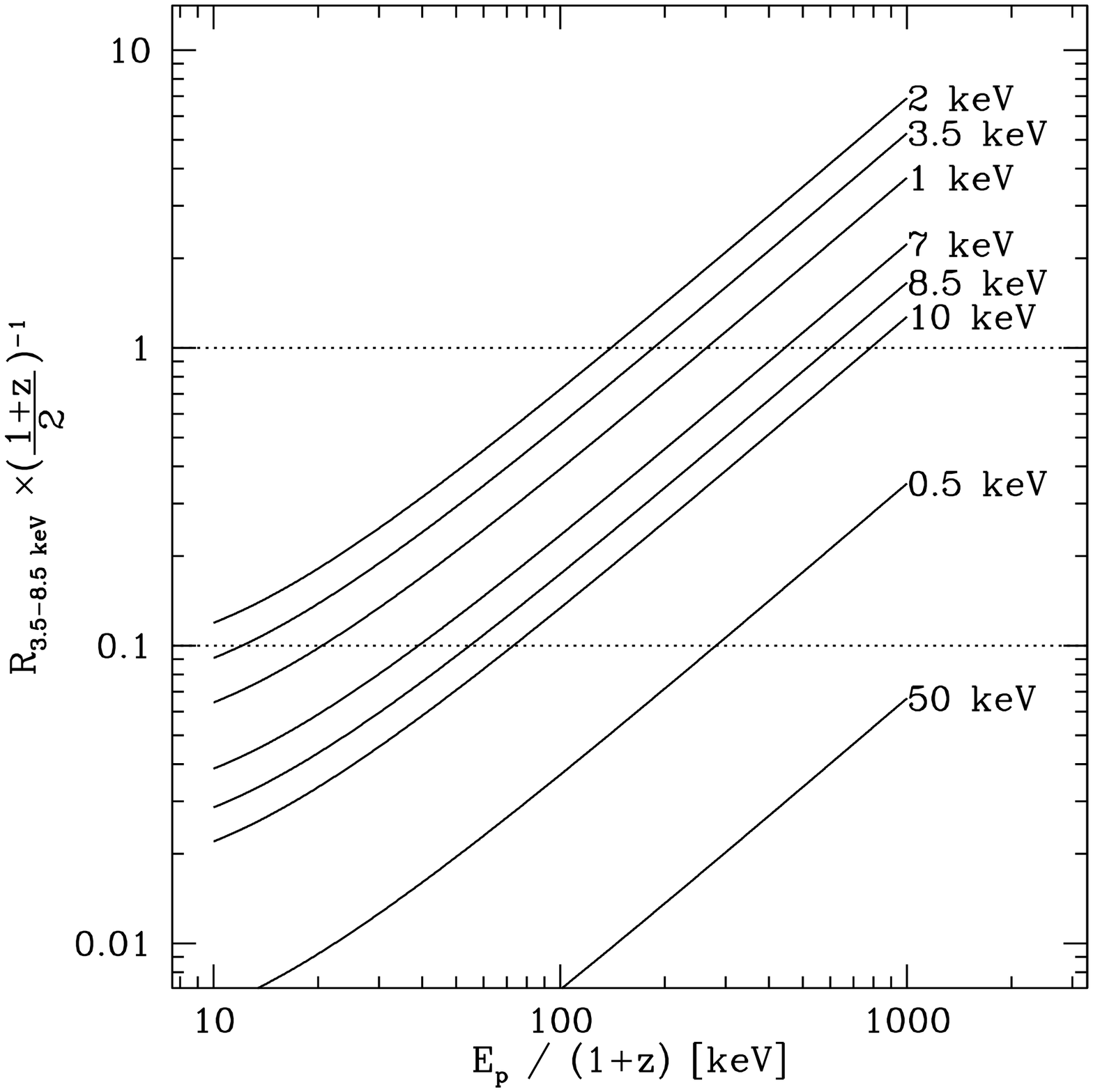}}
\resizebox{0.45\hsize}{!}{\includegraphics{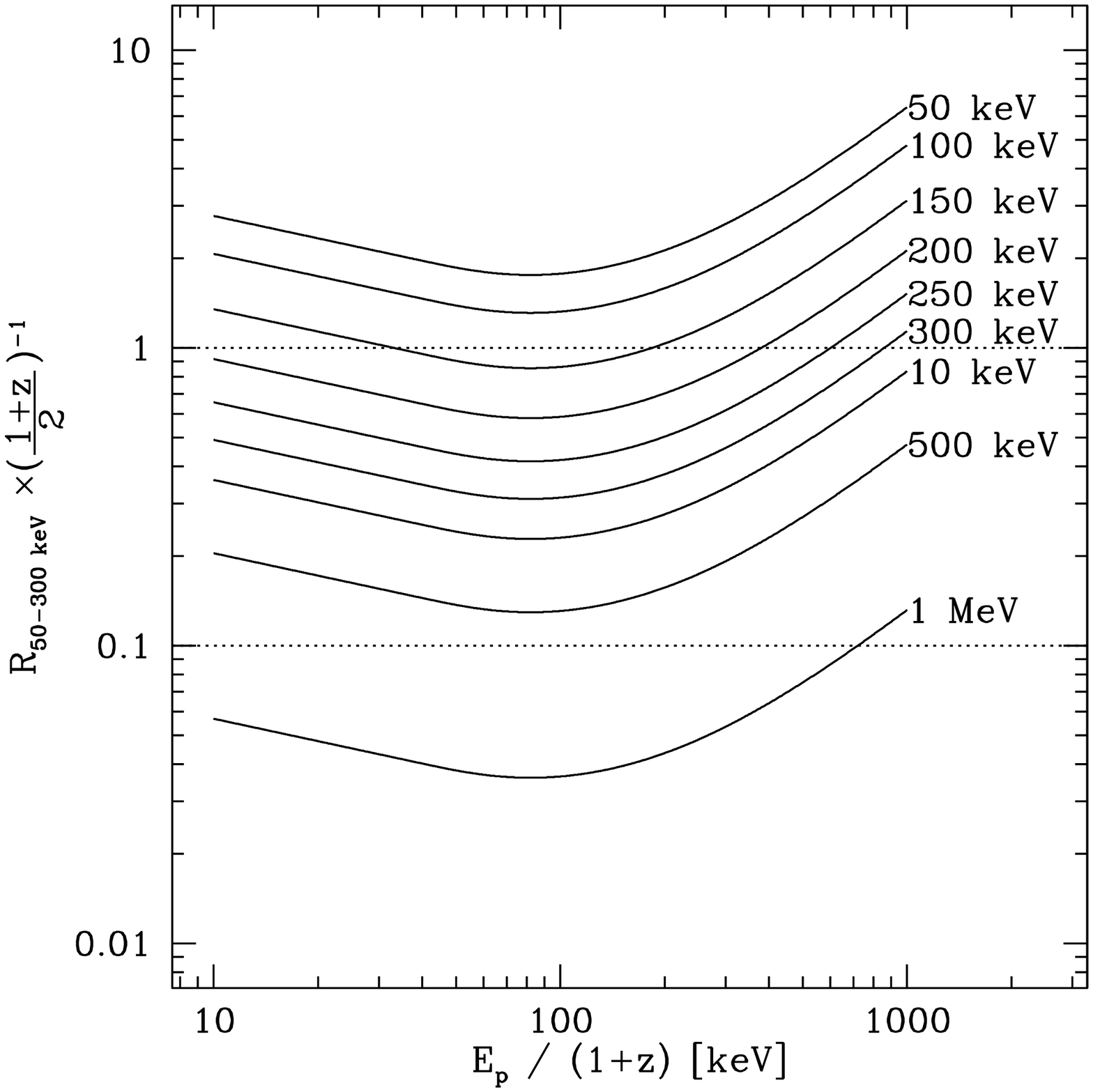}}
\caption{The ratio of the count rate due to the photospheric emission 
over the count rate due to the internal shocks is plotted as a
function of the peak energy of the non-thermal spectrum (observer
frame) for different values of the photospheric temperature $\left.k T_\mathrm{ph}\right./(1+z)$ in the observer frame.  The spectrum of the non-thermal emission from the internal shocks is
computed using the GRB-function with $\alpha=-1.0$ and $\beta=-2.25$. The following
parameters have been adopted : $\lambda=1$, $\dot{E}_{52}=1$, $\left.f_{\gamma}\right.=0.1$,
$\mu_{1}=1$. 
\textit{Left\,:} X-ray band $3.5$--$8.5\
\mathrm{keV}$; \textit{Right\,:} gamma-ray band $50$--$300\ \mathrm{keV}$.}
\label{fig:Ratio}
\end{figure*}


\section{Comparison with the emission from the internal
shocks}
\label{sec:Comparison}
\subsection{Time profile and spectrum of the emission from the
internal shocks}
We now estimate the count rate due to the emission of the internal shocks. Their  luminosity can be written as 
$\left.L_\mathrm{IS}\right. \simeq \left.f_{\gamma}\right. \dot{E}$,
with $\left.f_{\gamma}\right.\simeq f_\mathrm{d}\alpha_\mathrm{e}$. The efficiency $f_\mathrm{d}$ of the dissipation process is the
fraction of kinetic energy which is converted into internal energy
behind the shocks; $\alpha_\mathrm{e}$ is the fraction of the internal
energy which is injected into relativistic
electrons, which then radiate to produce the gamma-ray burst, with a
radiative efficiency $f_\mathrm{rad}$ which is assumed to be very close to
$1$. We do not discuss here the details of the radiative
processes 
and we simply assume that the emitted photons have a spectral distribution given by the ``GRB-function'' \citep{band:93}\,:
\begin{equation}
\frac{\mathrm{d}n^{\mathrm{IS}}(E)}{\mathrm{d}E\mathrm{d}t} =
\frac{1}{\mathcal{I}_{\mathrm{Band}}}\ \frac{\left.L_\mathrm{IS}\right.}{E_{\mathrm{p}}^{2}}\
\mathcal{B}\left(\frac{E}{E_{\mathrm{p}}}\right)\ .
\end{equation}
The peak energy $E_{\mathrm{p}}$ is defined as the maximum of $E^{2}\mathrm{d}n^{\mathrm{IS}}(E)/\mathrm{d}E/\mathrm{d}t$ and is measured here in the source frame, so that the peak energy in the observer frame is $E_{\mathrm{p}}/(1+z)$. The function $\mathcal{B}(x)$ has two parameters, the low and high energy slopes $\alpha$ and $\beta$ and is given by
\begin{equation}
\mathcal{B}(x) = \left\lbrace\begin{array}{ll}
x^{\alpha}\ \exp{\left(-\left(2+\alpha\right)x\right)} & \mathrm{if}\ x
\le \frac{\alpha-\beta}{2+\alpha}\\
x^{\beta} \left(\frac{\alpha-\beta}{2+\alpha}\right)^{\alpha-\beta}\
\exp{\left(\beta-\alpha\right)} & \mathrm{otherwise}
\end{array}\right.
\end{equation}
The integral $\mathcal{I}_{\mathrm{Band}}=\int_{0}^{+\infty}x\mathcal{B}(x)dx$ depends only on $\alpha$ and $\beta$ which we assume to be constant during the whole burst. The observed photon flux at energy $E$ is given by
\begin{equation}
C^{\mathrm{IS}}(E) = \frac{1}{\mathcal{I}_{\mathrm{Band}}}\ \frac{\left.L_\mathrm{IS}\right.}{4\pi D_{\mathrm{L}}^{2}}\ \left(\frac{1+z}{E_{\mathrm{p}}}\right)^{2}\ \mathcal{B}\left(\frac{(1+z)E}{E_{\mathrm{p}}}\right)
\end{equation}
and the corresponding count rate in the energy band $\left[E_{1};E_{2}\right]$ is
\begin{equation}
C_{12}^{\mathrm{IS}} =  2.5\ \frac{1+z}{D_{28}^{2}}
\left.f_{\gamma}\right._{0.1} \dot{E}_{52}
\left(\frac{E_{\mathrm{p}}}{200\ \mathrm{keV}}\right)^{-1}
\frac{\mathcal{I}^{\mathrm{IS}}_{12}}{\mathcal{I}_{\mathrm{Band}}}\ \mathrm{ph/cm^{2}/s},
\end{equation}
where $\mathcal{I}^{\mathrm{IS}}_{12}=\int_{x_{1}}^{x_{2}} \mathcal{B}(x) dx$ with $x_{1,2}=(1+z) E_{1,2}/E_{\mathrm{p}}$. 
\subsection{Comparison with the photospheric emission}
We now define $R_{12}$ as the ratio of the count rate due to the
photospheric emission over the count rate due to the internal shocks :
\begin{equation}
R_{12} = 1.6\ \lambda^{3/4}
\left.f_{\gamma}\right._{0.1}^{-1}
\dot{E}_{52}^{-1/4}\mu_{1}^{1/2} \frac{E_{\mathrm{p}}}{200\ \mathrm{keV}}\
 \frac{\mathcal{I}_{\mathrm{Band}}\,\mathcal{I}^{\mathrm{ph}}_{12}}{\mathcal{I}_{\mathrm{Planck}}\,\mathcal{I}^{\mathrm{IS}}_{12}}\ .
\label{eq:Ratio}
\end{equation}
Fig.~\ref{fig:Ratio} shows the value of $R_{12}$ for two energy bands
(X- and gamma-rays) as a function of $E_{\mathrm{p}}$ assuming different values
of the photospheric temperature $\left.k T_\mathrm{ph}\right.$. We have adopted $\alpha=-1.0$ and
$\beta=-2.25$ which are the typical slopes observed in GRBs
\citep{preece:00}. It is clear that the photospheric emission 
will show up
in a given band ($R_{12}\ga 0.1$) when the observed temperature $\left.k T_\mathrm{ph}\right./(1+z)$ of the photosphere crosses this band. 
These results show that with the prediction of
the standard fireball model for the photospheric temperature and
luminosity, it is very difficult to 
prevent the photospheric
emission 
from being easily detectable\footnote{Notice that $E_{\mathrm{p}}$ is defined as the peak energy of the non-thermal emission of the internal shocks and is of course no more the peak energy of the total observed spectrum when the photospheric emission is dominant.} either in the X-ray or gamma-ray range. 
The \mbox{presence} of a bright thermal component is not supported by the observations : the gamma-ray burst
prompt \mbox{emission}, as seen by \textit{BATSE}, is clearly 
non-thermal.
\mbox{Concerning} the X-ray emission, especially at
the beginning of the burst, more observations with better
spectro\-scopic capabilities than \textit{Beppo-SAX} will be necessary to check
wether a thermal component is present or not.
\begin{figure*}
\resizebox{0.45\hsize}{!}{\includegraphics{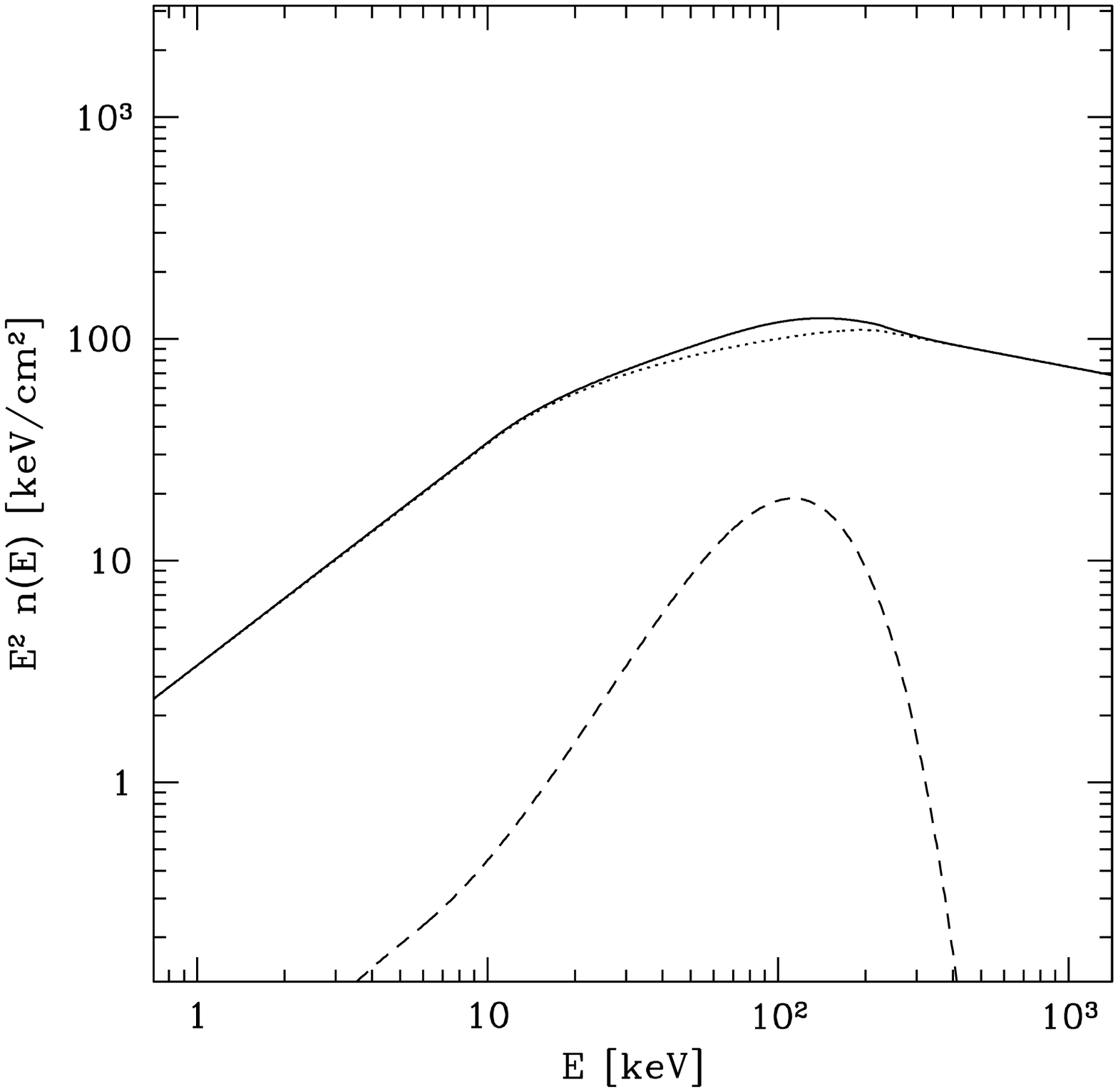}}
\resizebox{0.45\hsize}{!}{\includegraphics{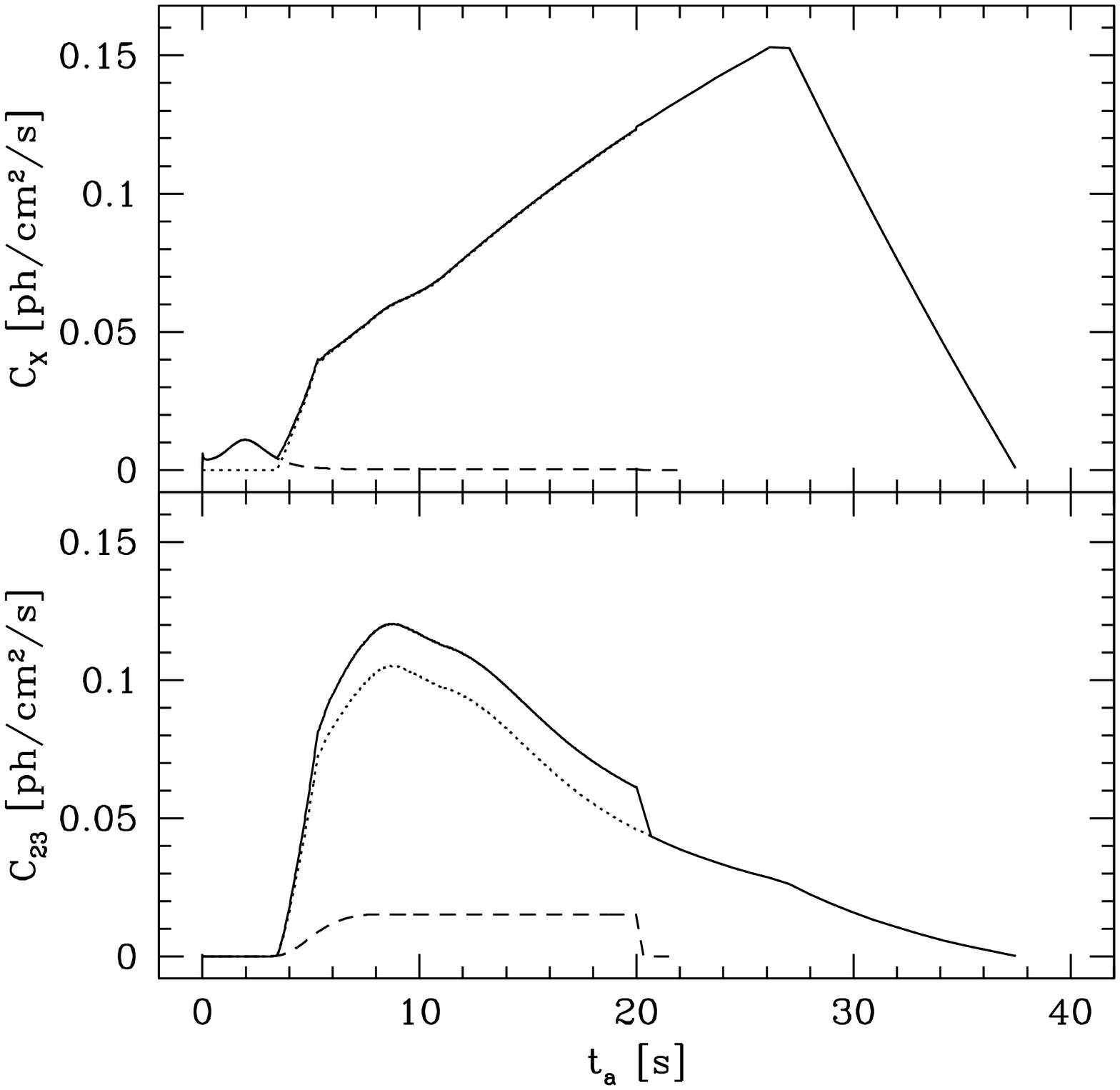}}
\caption{\textbf{A single pulse burst with a less luminous photosphere\,: spectrum and time profile}
(example considered in sec.~\ref{sec:ExampleC}).
The wind is the same as in sec.~\ref{sec:Example} but we now assume $\lambda=0.01$.
The redshift is $z=1$.
\textbf{Left\,:} the spectrum $E^{2}n(E)$ is plotted as a function of the photon energy in
$\mathrm{keV}$.  
\textbf{Right\,:} the count rate is plotted as a function of arrival time in the $3.5$--$8.5\ \mathrm{keV}$ band (top) and in the $50$--$300\
\mathrm{keV}$ band (bottom).  
As in fig.~\ref{fig:SpectrumProfiles}, 
the dashed line is the contribution of the photospheric emission, the dotted line is the contribution of
the internal shocks and the solid line is the total count rate (photosphere + internal shocks).}
\label{fig:SpectrumProfilesLow}
\end{figure*}

\subsection{Example : a single pulse burst}
\label{sec:ExampleC}
We have computed the emission of the internal shocks in the single pulse burst considered in secs.~\ref{sec:Example} and \ref{sec:ExamplePh} using the simple model developped by \citet{daigne:98}. 
The result is shown
in fig.~\ref{fig:SpectrumProfiles}. The equipartition
parameters have been chosen so that the peak energy of the emission from
the internal shocks is $200\ \mathrm{keV}$. As the photosphere
reaches a temperature $\left.k T_\mathrm{ph}\right./(1+z)$ as high as $94\ \mathrm{keV}$, the
study made in the previous subsection predicts that the photospheric
emission should be easily detectable, which is clearly visible in fig.~\ref{fig:SpectrumProfiles}.\\

To recover a burst which is dominated by the non-thermal emission of
the internal shocks in the gamma-ray range, eq.~\ref{eq:Ratio}
indicates that either $\left.f_{\gamma}\right.$ must increase or $\lambda$ must decrease. The first
solution is then to have more efficient internal shocks. In the example presented in fig.~\ref{fig:SpectrumProfiles}, the efficiency is low : $\left.f_{\gamma}\right. \simeq 0.02$.
As there are many uncertainities in the radiative processes leading to the observed gamma-ray emission, one could 
hope that $\left.f_{\gamma}\right.$ is indeed very close to $f_\mathrm{d}$, the fraction of kinetic energy dissipated in internal shocks. However, the efficiency $f_\mathrm{d}$ will never exceed a few $10^{-1}$. Therefore, even in the ideal case where $\left.f_{\gamma}\right.\sim f_\mathrm{d}$, one cannot expect to have $\left.f_{\gamma}\right.$ larger than $0.3$--$0.4$. We 
have checked 
 that this is not enough to avoid a 
easily
detectable photospheric emission.\\

The only remaining solution is then to assume that the photosphere is
less hot and luminous than what is predicted in the standard fireball
model, i.e. to decrease the value of $\lambda$ in eqs.~\ref{eq:Tphfint}--\ref{eq:Lphfint}.
 To have a spectrum entirely dominated by
the non-thermal component we have to adopt $\lambda \la 0.01$, which means
that less than 1 percent of the energy initially released by the
source is injected under internal energy form into a ``standard''
fireball. 
Such a situation is shown in fig.~\ref{fig:SpectrumProfilesLow} where we have computed the photospheric
emission of the 
single pulse burst
 with $\lambda=0.01$. It is impossible to detect the thermal emission of the
photosphere neither in the gamma-ray profile nor in the global
spectrum. However, in the X-ray range, one can notice that during 2
seconds before the internal shock emission starts, there is a thermal
precursor whose intensity is about 8 percents of the intensity at maximum
in the main pulse.\\

We have finally considered the effect of pair creation during the internal shock phase. The optical depth for pair creation is given by \citep{meszaros:00} :
\begin{equation}
\tau_{\pm} \simeq \frac{\alpha_\mathrm{\pm}\left.L_\mathrm{IS}\right.\sigma_\mathrm{T}}{4\pi r (m_\mathrm{e}c^{2}) \Gamma^{3} c}\ ,
\end{equation}
where $\alpha_\mathrm{\pm}$ is the fraction of the energy radiated in photons above the pair creation threshold.
Our internal shock model allows the computation of $\tau_{\pm}$ at each shock radius. For the example considered here, $\tau_{\pm}$ never exceeds $6\ 10^{-2}$. For larger $\dot{E}$ and / or smaller Lorentz factors (in this case, if $\dot{E}$ is increased by a factor of $\sim 20$ or if all Lorentz factors are divided by $\sim 2$), $\tau_{\pm}$ increases and pair creation can become important, especially for shocks occuring at small radii. We do not compute in this paper the detailed internal shock spectrum for this case where an additional thermal component can be expected.
\vspace*{-3ex}


\section{Discussion}
\label{sec:Discussion}
\subsection{X-ray thermal precursors}
\label{sec:XrayPrec}
In the GRBs observed by BeppoSAX the X-ray and gamma-ray
emission usually start simultaneously or the gamma-ray
emission starts earlier. 
Usually, no evidence is found for a thermal
component in the spectrum \citep{frontera:00}. 
Then, the
prompt X-ray emission is probably due to the internal
shocks like in the gamma-ray range. This implies
that the
photospheric emission must be present in these bursts only at a very low
level, i.e. $\lambda \ll 1$ as explained  in
the previous section. 
However, in at least one case -- GRB 990712 -- evidence was found in the spectrum for the presence during the burst of a weak thermal component of temperature 1.3 keV \citep{frontera:01}.  
In complement, a
 X-ray precursor activity has
been detected in a few GRBs by GINGA \citep{murakami:91} and
WATCH/GRANAT \citep{sazonov:98}. In the observations carried
out with the GRB detector onboard the GINGA satellite, X-ray
precursors were detected between 1.5 and 10 keV in about one third of the GRBs. The
spectrum of these X-ray precursors could be approximated by a
black-body  with temperatures between 1 and 2
keV. The WATCH catalog also includes several GRBs with X-ray
precursors detected between 8 and 20 keV.
As
can be seen in the time profiles of these bursts
\citep{sazonov:98}, the X-ray precursor usually has a
duration which is about 20-50\% 
of the duration of the 
whole burst
and its count rate in the 8-20 keV band 
reaches about 10-40\%
of the maximum count rate in the same band during the GRB.\\

\begin{figure}
\resizebox{\hsize}{!}{\includegraphics{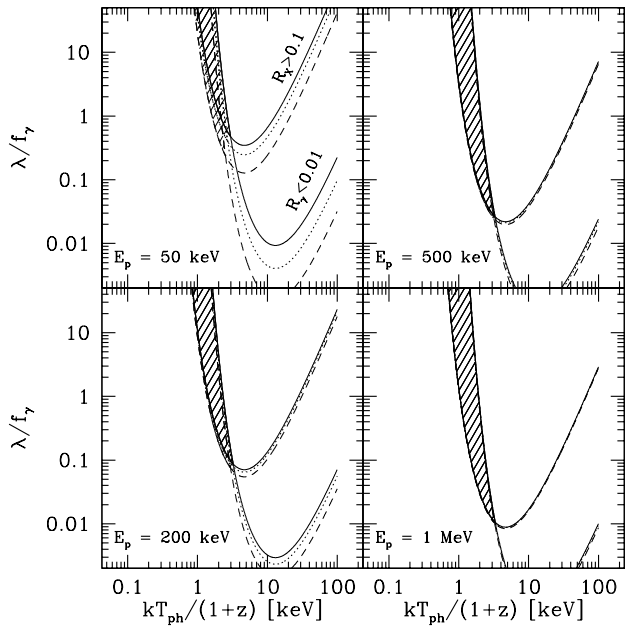}}
\caption{In the 
$\left.k T_\mathrm{ph}\right. / (1+z)$--$\lambda/\!\!\left.f_{\gamma}\right.$ plane, the stripped area shows the region
where the photospheric thermal emission is negligible in the 20-60 keV
gamma-ray band of WATCH/GRANAT (count rate ratio smaller than
1\%
) 
but can be detected in the 8-20 keV X-ray band of the
same experiment (count rate ratio greater than 10\%
). 
Each panel corresponds to a particular value of the peak energy $E_{\mathrm{p}}$
of
the internal shock emission (respectively 50, 200, 500 and 1000 keV).
The solid, dotted and dashed curves respectively correspond  to $z=0$, $z=1$ and $z=3$.
The other parameters are $\left.L_\mathrm{IS}\right.=\left.f_{\gamma}\right.\dot{E}=10^{51}\
\mathrm{erg/s}$, $\alpha=-1.$ and $\beta=-2.25$. }
\label{fig:XrayPrecursor}
\end{figure}

As the study of the GINGA data shows evidence for a thermal
origin, one can wonder whether these X-ray precursors 
are associated to the photospheric emission. 
This could be possible if the two following conditions are satisfied :\\
\noindent--\textit{Condition (1)} The ratio of the photospheric over
internal shock count rate as defined by eq.~\ref{eq:Ratio}
must be small in the gamma-ray range but greater than a few
10\% 
in the X-ray range. The region of the 
$\left.k T_\mathrm{ph}\right./(1+z)$--$\lambda / \left.f_{\gamma}\right.$ plane where such a condition can be
achieved (using the energy bands of the WATCH experiment) is
shown in fig.~\ref{fig:XrayPrecursor}. We find that (i) the
photospheric temperature $\left.k T_\mathrm{ph}\right./(1+z)$ must lie in the X-ray
band, which is easily obtained if $\lambda\simeq
0.1$. (ii) the ratio $\lambda/\left.f_{\gamma}\right.$ must be above a
minimal value which is typically about $0.1$ and decreases
when the peak energy $E_{\mathrm{p}}$ increases.\\
\noindent--\textit{Condition (2)} As no activity (thermal or non thermal) is detected
in the gamma-ray band during the X-ray precursor, the internal shock emission
must start at the end of the precursor. We have shown in
sec.~\ref{sec:ArrivalTime} that the arrival time of the
photons emitted by the layer ejected by the source at
$t_\mathrm{inj}$ when it becomes transparent can be approximated by
$t_\mathrm{a}\simeq t_\mathrm{inj}$. The arrival time of photons emitted by the
internal shocks due to the collisions between two layers
emitted at $t_\mathrm{inj}'$ and $t_\mathrm{inj}>t_\mathrm{inj}'$ (with
$\Gamma(t_\mathrm{inj})>\Gamma(t_\mathrm{inj}')$) is
\begin{equation}
t_\mathrm{a} \simeq
t_\mathrm{inj}+\frac{t_\mathrm{inj}-t_\mathrm{inj}'}{\left(\Gamma(t_\mathrm{inj}) / \Gamma(t_\mathrm{inj}')\right)^{2}-1}\ .
\end{equation}
The only possibility to increase the delay between the
beginning of the photospheric emission and the beginning of
the internal shock emission is then to impose that the
variability of the initial distribution of the Lorentz
factor in the relativitic wind is initially low
($\Gamma(t_\mathrm{inj})/\Gamma(t_\mathrm{inj}') \to 1$) and increases during
the wind production by the source.\\

\noindent Condition (1) is easily achieved if the initial
fraction of the energy released by the source under internal
energy form is low.
For instance, in
fig.~\ref{fig:SpectrumProfilesLow}, 
one clearly sees a X-ray precursor
lasting for about 10\% 
of the total duration with an
intensity of about 8\% 
of the intensity at maximum (in the
X-ray band). On the other hand, some of the precursors observed by
GINGA and WATCH/GRANAT have longer durations. This is
where condition (2), which is probably a stronger
constraint, is important.\\

We propose the following interpretation for the presence or absence of a precursor : it is necessary to have  $\lambda \ll 1$ in order to suppress a too strong thermal gamma-ray emission from the photosphere. This naturally leads to a prompt thermal X-ray activity, which then could be very frequent in GRBs. However this activity is too weak to be easily detected when it occurs simultaneously to the bright non-thermal emission from the internal shocks. It is only when it appears as a precursor activity that it can be clearly identified. This can happen if by chance the relativistic wind is initially produced with a smooth distribution
so that the internal shock activity is delayed. The expected features of such precursors are very close to the properties of the X-ray precursors observed by GINGA and WATCH/GRANAT.\\

To check the validity of this interpretation
one clearly needs
 more precise detections of the X-ray prompt
emission of GRBs and especially a better characterization of the
spectral properties of the X-ray precursors. If a
black-body spectrum can be identified without any ambiguity,
 the corresponding temperature will be measured, which would constrain the $\lambda$ parameter.

\vspace*{-1ex}

\subsection{The optical photospheric emission}
\label{sec:Optical}
To recover a dominant non-thermal gamma-ray emission 
we need $\lambda$ to be of a few percents or less.
The corresponding Planck spectrum then peaks
 in the X-ray band\,:
for instance, the burst considered
in sec.~\ref{sec:ExampleC} has a photospheric temperature
in the range $\left.k T_\mathrm{ph}\right. \simeq 2$--$100\ \mathrm{keV}$ for $\lambda=1$
and $\left.k T_\mathrm{ph}\right.\simeq 0.6$--$30\ \mathrm{keV}$ for $\lambda=0.01$. It
is interesting to estimate what is the photospheric
emission for even lower values of $\lambda$ and if it
could be dominant in the optical and  produce a
prompt optical flash comparable to that observed in GRB 990123 \citep{akerlof:99}. \\

It is very unlikely that the
photospheric emission peaks in the V band, because the
photospheric temperature scales as $\lambda^{1/4}$. To decrease
$\left.k T_\mathrm{ph}\right.$ from $100\ \mathrm{keV}$ to $1\ \mathrm{eV}$, a very
unrealistic value of $\lambda=(10^{-5})^{4}=10^{-20}$ is required
! The V band then always lies in the Rayleigh-Jeans part of
the photospheric spectrum where
$\frac{dn^\mathrm{ph}}{dEdt}\propto E$.
GRB 990123 has an averaged spectrum which is well reproduced by the
Band function with the following parameters\,: $\alpha=-0.6$,
$\beta=-3.11$, $E_{\mathrm{p}}=720\ \mathrm{keV}$ and photon flux $1.93\ 10^{-3}\ \mathrm{ph/s/cm^{2}/keV}$  at $1\ \mathrm{MeV}$
\citep{briggs:99}. The redshift of the source is
$z=1.6$.
The corresponding internal shock luminosity is
$\left.L_\mathrm{IS}\right.=\left.f_{\gamma}\right.\dot{E}\simeq 8.6\ 10^{52}\ \mathrm{erg/s}$. 
If
we now assume that the internal shocks have no other
contribution in the optical range than that given by the Band
spectrum, their flux in the V band ($0.55\ \mathrm{\mu m}$)
is
$F^\mathrm{IS}_\mathrm{V}\simeq 4.9\ 10^{-2}\ \mathrm{m Jy}$, 
which is much too low to explain the
optical flash reaching magnitude $m_\mathrm{V}\sim 9$ (i.e. 
$F_\mathrm{V}\simeq 0.92\ \mathrm{Jy}$ ) observed
by ROTSE. In the Rayleigh-Jeans regime, 
the corresponding flux due to the thermal photospheric emission is even lower :
\begin{equation}
F^\mathrm{ph}_\mathrm{V} \simeq 2.3\ 10^{-5}\ \left(\frac{\lambda}{\left.f_{\gamma}\right.}\right)^{3/4}\mu_{1}^{1/2}\left(\frac{\left.k T_\mathrm{ph}\right.}{1\ \mathrm{keV}}\right)^{-2}\ \mathrm{mJy}\ .
\end{equation}
We then find that the photospheric optical emission is much too weak to explain the ROTSE observations. This result is mainly due to the fact that the photospheric luminosity decreases much faster with $\lambda$ than the temperature.


\section{Conclusions}
\label{sec:Conclusions}
In the framework of the internal shock model for gamma-ray
bursts, we have computed in a detailed way the photospheric
emission of an ultra-relativistic wind with a variable
initial distribution of the Lorentz factor. We have compared
the obtained spectrum and time profile to the non-thermal
contribution of the internal shocks. Our main results are
the following :\\
\noindent \textit{(1) The photosphere in the standard fireball model is too hot and luminous.} In the standard fireball model where the initial
temperature of the fireball is about 1 MeV, the internal
energy is still large when the wind becomes transparent and
the photosphere is therefore hot and luminous. The
consequence is that the photospheric thermal component in the
X-ray/gamma-ray range is in most cases at least as bright as the
non-thermal component due to the internal shocks (even if
the internal shock efficiency is high). This is in
contradiction with the observations of BATSE and Beppo-SAX
showing non-thermal spectra.\\
\noindent \textit{(2) MHD winds are favored.} 
Results in much better agreement with the observations
are obtained when
it is assumed that only a small fraction $\lambda$ of the energy
released by the source is initially injected under
internal energy form in a fireball. Most of the energy could
for instance be initially under magnetic form, a large
fraction of the Poynting flux being eventually converted
into kinetic energy at large distances. For a typical internal shock efficiency of a few
percents, values of $\lambda \la 0.01$ are required, which means
that not more than 1\% 
of the energy is initially deposited in the ejected
matter (whose initial temperature is then of about a few
hundreds keV).\\
\noindent \textit{(3) X-ray thermal precursors can be obtained.} A consequence of this strong assumption is that moderately
low $\lambda$ ($\lambda \simeq $ a few percents) lead to the
presence of thermal X-ray precursors if the
distribution of the Lorentz factor is not too variable in the
initial phase of wind production. The characteristics of
these precursors (spectral range, duration,
intensity) are very comparable to the X-ray precursor
activity observed in several GRBs by GINGA and WATCH/GRANAT.\\
\noindent \textit{(4) The optical photospheric emission is very weak.} 
For very small $\lambda$ values,
the photospheric emission can be shifted to even lower energies. 
However, we have
shown that it also becomes much too weak to explain the
prompt optical emission observed by ROTSE in GRB 990123.\\

A good test of the results presented in this paper 
would be the detection of X-ray precursors
by an instrument with good spectral capabilities, so that
a thermal origin could be firmly established. A determination of the photospheric temperature
would put an interesting constraint on the
$\lambda/\!\!\left.f_{\gamma}\right.$ ratio and then on the wind acceleration mechanism. Moreover, if the photospheric thermal
emission could be clearly detected (for instance in the soft
X-ray range), it would provide a direct information about the initial distribution of the Lorentz factor in
the wind before the internal shocks start.\\
\vspace*{-3ex}

\section*{Acknoledgments}
F.D. acknowledges financial support from a postdoctoral fellowship from the 
French Spatial Agency (CNES).

\bibliographystyle{mn2e}
\bibliography{grbprec}

\end{document}